\documentclass[aip,jap,amsmath,amssymb,preprint,eqsecnum]{revtex4-1}
\usepackage{graphicx}
\usepackage{bm}
\usepackage{upgreek}

\begin{document}

\preprint{JAP}

\title{Eddy current effects in the magnetization dynamics of ferromagnetic metal nanoparticles}
\author{S.~I.~Denisov}
\email{denisov@sumdu.edu.ua}
\author{T.~V.~Lyutyy}
\author{B.~O.~Pedchenko}
\author{H.~V.~Babych}
\affiliation{Sumy State University, 2 Rimsky-Korsakov Street, UA-40007 Sumy, Ukraine}


\begin{abstract}
 We develop an analytical model for describing the magnetization dynamics in ferromagnetic metal nanoparticles, which is based on the coupled system of the Landau-Lifshitz-Gilbert (LLG) and Maxwell equations. By solving Maxwell's equations in the quasi-static approximation and finding the magnetic field of eddy currents, we derive the closed LLG equation for the magnetization that fully accounts for the effects of conductivity. We analyze the difference between the LLG equations in metallic and dielectric nanoparticles and show that these effects can strongly influence the magnetization dynamics. As an example illustrating the importance of eddy currents, the phenomenon of precessional switching of magnetization is considered.
\end{abstract}



\maketitle

\section{INTRODUCTION}
\label{Intr}

The magnetization dynamics in ferromagnetic materials with strong exchange interaction between atomic magnetic moments is well described by the phenomenological Landau-Lifshitz (LL) or Landau-Lifshitz-Gilbert (LLG) equation.\cite{LaLi1, Gilb} Although the damping term in the LLG equation is physically more relevant than the corresponding term in the LL equation,\cite{Kiku} these equations are mathematically equivalent and equally efficient. According to them, the local magnetization undergoes damped precessional motion about the local effective magnetic field, which usually includes the exchange and anisotropy fields, the externally applied magnetic field and the magnetostatic field. Except in the simplest cases (see, e.g., Refs.~[\onlinecite{ KIK, Lak, BBI}] and references therein), the LLG equation (because of the equivalence of the LL and LLG equations, we refer only to the LLG one) can not be solved exactly. Moreover, since the magnetostatic field depends on the magnetization distribution, the LLG equation must be solved together with the magnetostatic Maxwell equations, i.e., the LLG equation is in general not closed. In metallic materials, the effective field includes also the magnetic field of eddy currents that are induced by changing in time the external magnetic field and the magnetization direction. As a consequence, the LLG equation must be supplemented by the full system of Maxwell's equations. This coupled system of the LLG and Maxwell equations can be solved numerically using, e.g., advanced methods discussed in Refs.~[\onlinecite{SlBa, Cimr, LeTr}].

It is well known\cite{Brow1} that if the ferromagnetic particles are sufficiently small (with the particles size of the order of $10^{2}$ nanometers or less) then the uniform magnetization is energetically preferable. These nanoparticles exhibit unique properties and have  many current and potential applications, e.g., in data storage,\cite{Ross, MTMA, Kiki} spintronics\cite{WABD, ZFS} and biomedicine.\cite{PCJD, Ferr, LaLe, LFPR} If, in addition, the nanoparticles are ellipsoidal, when the internal magnetostatic field is strictly uniform,\cite{Brow2} or if in non-ellipsoidal nanoparticles this field is assumed to be uniform and known, then the LLG equation becomes closed. This equation, which due to the uniform magnetization reduces to the first order vector differential equation, is a valuable tool for studying the magnetization dynamics in dielectric nanostructures. In particular, it was used to study the periodic and quasiperiodic regimes of the magnetization precession,\cite{BSM, BMS1, DLHT, BMS} chaotic magnetization dynamics,\cite{APC, VaPo, BPSV, LBGP} precessional magnetization switching,\cite{BWH, BFH, KaRu, SMB, SuWa} thermal\cite{Brow3, GaLa, CKW, DLH, DPL} and many other effects.

In conducting nanostructures the LLG equation becomes unclosed because of the presence of the magnetic field of eddy currents. Stimulated by potential applications, the temporal evolution of the (non-uniform) magnetization and eddy currents in some nanostructures has been studied by numerically solving the coupled system of the LLG and Maxwell equations. In particular, this approach was used to analyze the process of magnetization reversal in metallic nanocubes.\cite{ToLo, ToMa, HrKi} It was shown that, similarly to the case of domain walls moving in metallic ferromagnets, eddy currents act on the reversal process as an additional damping parameter in the LLG equation. Arising from Faraday's law of induction, this result seems to be quite general. However, even in the simplest case of a uniformly magnetized spherical nanoparticle, considered in Ref.~[\onlinecite{MaLo}], the problem of reducing the coupled system of the LLG and Maxwell equations to the closed LLG equation has not been solved completely. In particular, the authors determined the magnetic field of eddy currents only in the nanoparticle centre and used it to find the contribution of eddy currents to the damping parameter. But because the current-induced magnetic field inside the nanoparticle is strongly non-uniform, this contribution differs significantly from the exact one (see below).

In this paper, we consider a more general model in which the nanoparticle and its environment have different magnetic susceptibilities, provide an exact analytical solution of the quasi-static Maxwell's equations, and analyze in detail the role of eddy currents in the magnetization dynamics. The paper is organized as follows. In Sec.~\ref{Model}, we describe the model and introduce the coupled system of the LLG and Maxwell equations. By solving Maxwell's equations in the quasi-static approximation, we determine the induced electric field and the magnetic field of eddy currents in Secs.~\ref{Electr} and \ref{Magn}, respectively. The closed LLG equation, which accounts for the effects of eddy currents in the magnetization dynamics, is derived in Sec.~\ref{Eff}. In the same section, we demonstrate the importance of eddy currents for the correct description of precessional switching of magnetization in metallic nanoparticles. Finally, our findings are summarized in Sec.~\ref{Concl}.

\section{MODEL AND BASIC EQUATIONS}
\label{Model}

We consider a single-domain ferromagnetic particle of radius $a$ which is characterized by the electric conductivity $\sigma$ and magnetic susceptibility $\mu_{1}$. It is also assumed that the particle is electrically neutral and is embedded in a dielectric matrix, whose magnetic susceptibility equals $\mu_{2}$, and the origin of the Cartesian coordinate system $xyz$ is assumed to be located at the centre of the particle. If the exchange energy between neighboring spins significantly exceeds the magnetic energy $W$ of the particle, then the particle magnetization $\mathbf{M} = \mathbf{M}(t)$ changes with time in such a way that $|\mathbf{M}| = M = \mathrm{const}$. In this case, the dynamics of $\mathbf{M}$ can be described by the LLG equation\cite{Gilb}
\begin{equation}
    \dot{\mathbf{M}} = -\gamma \mathbf{M} \times
    \bm{\mathcal{H}}_{\mathrm{eff}}
    + \frac{\alpha}{M}\, \mathbf{M} \times
    \dot{\mathbf{M}}.
    \label{LLG}
\end{equation}
Here, $\gamma(>0)$ is the gyromagnetic ratio, $\alpha(>0)$ is the Gilbert damping parameter, the overdot denotes differentiation with respect to time, the cross stands for the cross (vector) product, and $\bm{\mathcal{H}}_{\mathrm{eff}}$ is the total effective magnetic field acting on the magnetization vector.

Due to the particle conductivity, it is convenient to represent the total effective field as a sum of two terms: $\bm{\mathcal{H}}_{\mathrm{eff}} = \mathbf{H}_{\mathrm{eff}} + \overline{\mathbf{H}}$, where $\mathbf{H}_{ \mathrm{ eff}} = -(1/V) \partial W/ \partial \mathbf{M}$ is the effective magnetic field at $\sigma=0$ and $\overline{\mathbf{H}} = (1/V) \int_{V} \mathbf{H} d\mathbf{r}$ is the averaged (over the particle volume $V=4\pi a^{3}/3$) magnetic field of eddy currents. In particular, if the particle is magnetically uniaxial then
\begin{equation}
    \mathbf{H}_{\mathrm{eff}} = \frac{H_{a}}{M}(
    \mathbf{M} \cdot \mathbf{e}_{a})\mathbf{e}_{a}
    + \mathbf{H}_{1}^{(0)}.
    \label{H_eff}
\end{equation}
Here, $H_{a}$ is the anisotropy field, $\mathbf{e}_{a}$ is the unit vector along the anisotropy axis, the dot denotes the dot (scalar) product, and $\mathbf{H }_{1} ^{(0)}$ is the magnetic field inside the particle. By solving the magnetostatic equations for a given geometry, it can be easily shown\cite{BaTo} that
\begin{equation}
    \mathbf{H}^{(0)}_{1} = \kappa\mathbf{H}_{0} -
    \frac{4\pi \kappa}{3\mu_{2}}\mathbf{M}
    \label{H^0_1}
\end{equation}
with
\begin{equation}
    \kappa = \frac{3\mu_{2}}{\mu_{1} + 2\mu_{2}}.
    \label{kappa}
\end{equation}
The first term on the right-hand side of Eq.~(\ref{H^0_1}) is the uniform  magnetic field induced by the external magnetic field $\mathbf{H}_{0} = \mathbf{H}_{0}(t)$, and the second term represents the demagnetization field induced by the magnetization.

An important feature of Eq.~(\ref{LLG}) is that it is not closed. This is because the magnetic field $\mathbf{H} = \mathbf{H} (\mathbf{r},t)$ of eddy currents itself depends on the magnetization $\mathbf{M}$. Therefore, Eq.~(\ref{LLG}) in the case of conducting particles must be solved together with the Maxwell equations. In the quasi-static approximation, these equations (in CGS units) can be written as follows\cite{LaLi2}:
\begin{subequations}\label{EH}
\begin{eqnarray}
    &\displaystyle \nabla \times \mathbf{E}_{l}
    = -\frac{1}{c} \frac{\partial}{\partial t}
    \mathbf{B}_{l}, \qquad
    \nabla \cdot \mathbf{E}_{l} = 0,
    \label{eqs_E_l}
    \\[4pt]
    &\displaystyle \nabla \times \mathbf{H}_{l}
    = \frac{4\pi \sigma_{l}}{c} \mathbf{E}_{l},
    \qquad \nabla \cdot \mathbf{H}_{l} = 0.
    \label{eqs_H_l}
\end{eqnarray}
\end{subequations}
Here, $\mathbf{E}_{l} = \mathbf{E}_{l}(\mathbf{r},t)$ is the induced electric field, the indexes $l=1$ and $l=2$ denote the corresponding quantities inside ($r = |\mathbf{r}| <a$) and outside ($r>a$) the particle, respectively, $\sigma_{l} = \sigma \delta_{1l}$, $\delta_{1l}$ is the Kronecker delta, $c$ is the light speed in vacuum, $\nabla \times$ is the curl, $\nabla \cdot$ is the divergence, and $\mathbf{B}_{l} = \mathbf{B}_{l}(\mathbf{r},t)$ is the magnetic induction. In addition to the conditions under which Eqs.~(\ref{EH}) hold, we use also the condition (for details, see Sec.~\ref{Eff}) that permits us to use the magnetostatic approximation for finding $\mathbf{B}_{1}$ and $\mathbf{B}_{2}$. According to Ref.~\onlinecite{BaTo}, in this approximation $\mathbf{B}_{1} = \mu_{1} \mathbf{H}^{(0)}_{1} + 4\pi \mathbf{M}$ and $\mathbf{B}_{2} = \mu_{2} \mathbf{H}^{(0)}_{2}$, where
\begin{equation}
    \mathbf{H}^{(0)}_{2} = \mathbf{H}_{0} -
    \frac{\mathbf{m}^{(0)}}{r^{3}} +
    \frac{3(\mathbf{m}^{(0)} \cdot
    \mathbf{r}) \mathbf{r}}{r^{5}}
    \label{H^0_2}
\end{equation}
represents the magnetic field outside the particle and
\begin{equation}
    \mathbf{m}^{(0)} = \frac{a^{3} \kappa}{3\mu_{2}}
    [(\mu_{1} - \mu_{2})\mathbf{H}_{0} + 4\pi \mathbf{M}]
    \label{m^0}
\end{equation}
is the particle magnetic moment induced by the external magnetic field and magnetization. Note also that the tangential and normal components of the vectors $\mathbf{E}_{l}$ and $\mathbf{H}_{l}$ (denoted by the indices $\tau$ and $n$, respectively) must satisfy the following boundary conditions:
\begin{subequations}\label{bc_EH}
\begin{eqnarray}
    &\mathbf{E}_{1\tau} = \mathbf{E}_{2\tau}, \quad
    E_{2n} = 0,
    \label{bc_E}
    \\[4pt]
    &\mathbf{H}_{1\tau} = \mathbf{H}_{2\tau}, \quad
    \mu_{1}H_{1n} = \mu_{2}H_{2n}.
    \label{bc_H}
\end{eqnarray}
\end{subequations}

\section{INDUCED ELECTRIC FIELD}
\label{Electr}

According to Eq.~(\ref{H^0_1}), the magnetic induction $\mathbf{B}_{1}$ in the magnetostatic approximation reads
\begin{equation}
    \mathbf{B}_{1} = \kappa\Big(\mu_{1}\mathbf{H}_{0}
    + \frac{8\pi}{3} \mathbf{M} \Big).
    \label{B1}
\end{equation}
Because $\mathbf{B}_{1}$ does not depend on $\mathbf{r}$, we seek the induced electric field $\mathbf{E}_{1}$ in the form $\mathbf{E}_{1} = \mathbf{a}(t) \times \mathbf{r}$. For this representation of $\mathbf{E}_{1}$, the second equation in (\ref{eqs_E_l}) holds identically and the first one yields
\begin{equation}
    \mathbf{E}_{1} = -\frac{\kappa}{2c} \Big( \mu_{1}
    \dot{\mathbf{H}}_{0} + \frac{8\pi}{3}
    \dot{\mathbf{M}} \Big)\! \times \mathbf{r}.
    \label{E1}
\end{equation}
It is this electric field which induces eddy currents of density $\mathbf{J} = \sigma \mathbf{E}_{1}$ inside the particle.

Similarly, from Eqs.~(\ref{H^0_2}) and (\ref{m^0}), for the magnetic induction outside the particle we obtain
\begin{eqnarray}
    \mathbf{B}_{2}  \!\!&=&\!\! \mu_{2}\mathbf{H}_{0}
    - \frac{(\mu_{1} - \mu_{2})\kappa V}{4\pi r^{5}}
    [r^{2}\mathbf{H}_{0} - 3(\mathbf{H}_{0}\cdot
    \mathbf{r})\mathbf{r}]
    \nonumber\\ [4pt]
    &&\!\! - \frac{\kappa V}{r^{5}} [r^{2}\mathbf{M}
    - 3(\mathbf{M}\cdot \mathbf{r})\mathbf{r}].
    \label{B2}
\end{eqnarray}
This suggests to seek the induced electric field at $r>a$ in the form
\begin{equation}
    \mathbf{E}_{2} = [u(r)\dot{\mathbf{H}}_{0}
    + v(r)\dot{\mathbf{M}}] \times \mathbf{r},
    \label{E2a}
\end{equation}
where the functions $u(r)$ and $v(r)$ should be determined. As before, equation $\nabla \cdot \mathbf{E}_{2} = 0$ is satisfied identically and, in accordance with Eq.~(\ref{E2a}), the curl of $\mathbf{E}_{2}$ is given by
\begin{eqnarray}
    \nabla \times \mathbf{E}_{2} \!\!&=&\!\! [ru'(r)
    + 2u(r)]\dot{\mathbf{H}}_{0} - \frac{u'(r)}{r}
    (\dot{\mathbf{H}}_{0} \cdot \mathbf{r})\mathbf{r}
    \nonumber\\ [4pt]
    &&\!\! + [rv'(r) + 2v(r)]\dot{\mathbf{M}} -
    \frac{v'(r)}{r}(\dot{\mathbf{M}} \cdot \mathbf{r})
    \mathbf{r}. \quad
    \label{curlE2}
\end{eqnarray}
Then, using Eq.~(\ref{B2}) and equating the right-hand sides of Eq.~(\ref{curlE2}) and equation $\nabla \times \mathbf{E}_{2} = - (1/c) \dot{ \mathbf{B}}_{2}$, we make sure that $u(r)$ must satisfy the equations
\begin{eqnarray}
    &\displaystyle ru'(r) + 2u(r) = -\frac{\mu_{2}}{c}
    + \frac{(\mu_{1} - \mu_{2})\kappa V}{4\pi cr^{3}}, &
    \nonumber\\ [4pt]
    &\displaystyle u'(r) = \frac{3(\mu_{1} - \mu_{2})
    \kappa V}{4\pi cr^{4}}&
    \label{eq_u}
\end{eqnarray}
and $v(r)$ the equations
\begin{equation}
    rv'(r) + 2v(r) = \frac{\kappa V}{cr^{3}},
    \quad v'(r) = \frac{3\kappa V}{cr^{4}}
    \label{eq_v}
\end{equation}
(the prime denotes differentiation with respect to $r$). These equations are easily solved,
\begin{equation}
    u(r) = -\frac{\mu_{2}}{2c} -\frac{(\mu_{1}
    - \mu_{2})\kappa V}{4\pi cr^{3}},
    \quad v(r) = -\frac{\kappa V}{cr^{3}},
    \label{u,v}
\end{equation}
and from Eq.~(\ref{E2a}) one finds the induced electric field outside the particle
\begin{equation}
    \mathbf{E}_{2} = - \frac{\kappa}{2c}
    \Big[\Big( \frac{\mu_{2}} {\kappa} +
    \frac{(\mu_{1} - \mu_{2})V}{2\pi r^{3}}
    \Big) \dot{\mathbf{H}}_{0} + \frac{2V}
    {r^{3}} \dot{\mathbf{M}}\Big]\! \times
    \mathbf{r}.
    \label{E2}
\end{equation}

Using Eqs.~(\ref{E1}) and (\ref{E2}), it is not difficult to verify that the boundary conditions (\ref{bc_E}) are identically fulfilled. It should also be emphasized that since the quasi-static approximation is used, Eq.~(\ref{E2}) correctly describes the induced electric field at distances not too far from the particle surface. But, as follows from Eq.~(\ref{eqs_H_l}) (see also below), this fact does not affect the magnetic field of eddy currents both inside and outside the particle.

\section{MAGNETIC FIELD OF EDDY CURRENTS}
\label{Magn}

The induced electric field (\ref{E1}) shows that the magnetic field of eddy currents can be represented in the form
\begin{equation}
    \mathbf{H}_{l} = f_{l}(r)\dot{\mathbf{H}}_{0} +
    g_{l}(r)(\dot{\mathbf{H}}_{0}\cdot \mathbf{r})
    \mathbf{r} + p_{l}(r)\dot{\mathbf{M}} + q_{l}(r)
    (\dot{\mathbf{M}}\cdot \mathbf{r})\mathbf{r}
    \label{Hl}
\end{equation}
with so far unknown functions $f_{l}(r)$, $g_{l}(r)$, $p_{l}(r)$, and $q_{l}(r)$. From this it follows straightforwardly that
\begin{eqnarray}
    \nabla \times \mathbf{H}_{l} \!\!&=&\!\!
    -[f'_{l}(r)/r - g_{l}(r)] \dot{\mathbf{H}}_{0}
    \times \mathbf{r}
    \nonumber\\ [4pt]
    &&\!\! -[p'_{l}(r)/r - q_{l}(r)]\dot{\mathbf{M}}
    \times \mathbf{r}
    \label{curlHl}
\end{eqnarray}
and
\begin{eqnarray}
    \nabla \cdot \mathbf{H}_{l} \!\!&=&\!\!
    [f'_{l}(r)/r + rg'_{l}(r) + 4g_{l}(r)]
    \dot{\mathbf{H}}_{0} \cdot \mathbf{r}
    \nonumber\\ [4pt]
    &&\!\! + [p'_{l}(r)/r + rq'_{l}(r) +
    4q_{l}(r)]\dot{\mathbf{M}} \cdot
    \mathbf{r}.
    \label{divHl}
\end{eqnarray}
Using these formulas, the first equation in (\ref{eqs_H_l}) yields
\begin{eqnarray}
    &\displaystyle f'_{l}(r) - rg_{l}(r) = \frac{2\pi
    \sigma \kappa\mu_{1}}{c^{2}}r \delta_{1l},&
    \nonumber\\ [4pt]
    &\displaystyle p'_{l}(r) - rq_{l}(r) = \frac{16
    \pi^{2} \sigma \kappa}{3c^{2}}r \delta_{1l}&
    \label{eqs1}
\end{eqnarray}
and the second equation in (\ref{eqs_H_l}) reduces to
\begin{eqnarray}
    &\displaystyle f'_{l}(r) + r^{2}g'_{l}(r) +
    4rg_{l}(r) = 0,&
    \nonumber\\ [6pt]
    &\displaystyle p'_{l}(r) + r^{2}q'_{l}(r) +
    4rq_{l}(r) = 0.&
    \label{eqs2}
\end{eqnarray}

Considering the region inside the particle (when $l=1$) and assuming that the physically reasonable condition $|\mathbf{H}_{1}| < \infty$ holds, from Eqs.~(\ref{eqs1}) and (\ref{eqs2}) one obtains
\begin{eqnarray}
    &\displaystyle f_{1}(r) = \phi - 2r^{2}g_{1}(r),
    \quad g_{1}(r) = - \frac{3\tau_{\sigma}}{2\kappa a^{2}},&
    \nonumber\\ [4pt]
    &\displaystyle p_{1}(r) = \psi - 2r^{2}q_{1}(r),
    \quad q_{1}(r) = - \frac{4\pi \tau_{\sigma}}{\kappa
    \mu_{1} a^{2}},&\quad
    \label{fgpq1}
\end{eqnarray}
where $\phi$ and $\psi$ are constants of integration and
\begin{equation}
    \tau_{\sigma} = \frac{4\pi \sigma \kappa^{2}
    a^{2}\mu_{1}}{15c^{2}}
    \label{def_tau}
\end{equation}
is the characteristic time. It is not difficult to verify that outside the particle (when $l=2$) the solution of Eqs.~(\ref{eqs1}) and (\ref{eqs2}), which satisfies the natural condition $|\mathbf{H} _{2}| \to 0$ as $r \to \infty$, is given by
\begin{eqnarray}
    &\displaystyle f_{2}(r) = -\frac{\nu}{3r^{3}},
    \quad g_{2}(r) = \frac{\nu}{r^{5}},&
    \nonumber\\ [4pt]
    &\displaystyle p_{2}(r) = -\frac{\epsilon}{3r^{3}},
    \quad q_{2}(r) = \frac{\epsilon}{r^{5}}.&
    \label{fgpq2}
\end{eqnarray}

To determine the constants of integration $\phi$, $\psi$, $\nu$ and $\epsilon$, we use the boundary conditions (\ref{bc_H}). Taking into account that $\mathbf{H}_{1,2 \tau} = \mathbf{e}_{n} \times (\mathbf{H}_{1,2} \times \mathbf{e}_{n})|_{r=a}$ and $H_{1,2 n} = \mathbf{H}_{1,2} \cdot \mathbf{e}_{n} |_{r=a}$ ($\mathbf{e}_{n} = \mathbf{r}/r$), these boundary conditions together with Eqs.~(\ref{Hl}), (\ref{fgpq1}) and (\ref{fgpq2}) lead to the following system of algebraic equations:
\begin{eqnarray}
    &\displaystyle \phi + \frac{\nu}{3a^{3}} = -
    \frac{3\tau_{ \sigma}}{\kappa}, \quad \psi +
    \frac{\epsilon}{3a^{3}} = - \frac{4\pi \tau_{
    \sigma}} {\kappa \mu_{1}},&
    \nonumber\\ [4pt]
    &\displaystyle \phi - \frac{2\mu_{2}\nu}{3
    \mu_{1}a^{3}} = - \frac{3\tau_{\sigma}}
    {2\kappa}, \quad  \psi - \frac{2\mu_{2}
    \epsilon}{3\mu_{1}a^{3}} = - \frac{4\pi
    \tau_{\sigma}}{\kappa \mu_{1}}.&\quad
    \label{syst}
\end{eqnarray}
Solving it with respect to the mentioned constants of integration, one gets
\begin{eqnarray}
    &\displaystyle \phi = -\Big(1 + \frac{3}
    {2\kappa}\Big)\tau_{\sigma}, \quad
    \nu = - \frac{3\mu_{1}}{2\mu_{2}} a^{3}
    \tau_{\sigma},&
    \nonumber\\ [4pt]
    &\displaystyle \psi =  - \frac{8\pi}{3\mu_{1}}
    \Big(1 + \frac{3}{2\kappa}\Big)\tau_{\sigma},
    \quad \epsilon = - \frac{4\pi}{\mu_{2}}
    a^{3}\tau_{\sigma}.&\quad
    \label{par}
\end{eqnarray}

\begin{figure}
    \centering
    \includegraphics[totalheight=6cm]{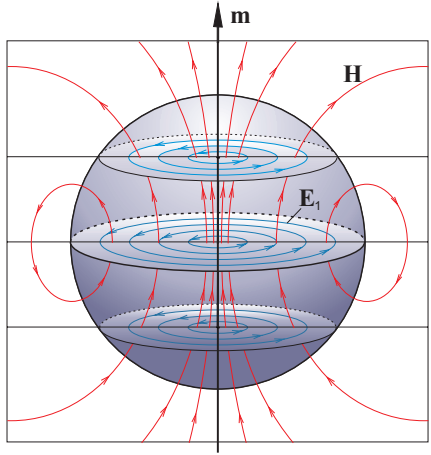}
    \caption{\label{fig1} (Color online) Plots of the electric
    and magnetic field lines for $\mu_{1} = \mu_{2} = 1$. Since
    according to Eqs.~(\ref{E1}) and (\ref{m}) $\mathbf{E}_{1}
    = (\kappa \mu_{2}/ca^{3} \tau_{\sigma})\mathbf{m} \times
    \mathbf{r}$, the lines of the induced electric field are
    circular and lie in planes perpendicular to the vector
    $\mathbf{m}$. The lines of the magnetic field $\mathbf{H}$
    of eddy currents, determined by Eqs.~(\ref{H1}) and
    (\ref{H2}), are shown in the plane of the figure.}
\end{figure}

Thus, collecting the above results, we find the magnetic field of eddy currents inside the particle
\begin{equation}
    \mathbf{H}_{1} = \frac{\mu_{2}}{\kappa
    \mu_{1}}\Big[\Big(3 + 2\kappa - 6\frac{r^{2}}
    {a^{2}} \Big)\frac{\mathbf{m}}{a^{3}}
    + \frac{3}{a^{5}}(\mathbf{m}\cdot
    \mathbf{r})\mathbf{r}\Big]
    \label{H1}
\end{equation}
and outside the particle
\begin{equation}
    \mathbf{H}_{2} = -\frac{\mathbf{m}}{r^{3}}
    + \frac{3}{r^{5}}(\mathbf{m}\cdot \mathbf{r})
    \mathbf{r},
    \label{H2}
\end{equation}
where
\begin{equation}
    \mathbf{m} = -\frac{a^{3} \tau_{\sigma}}
    {2\mu_{2}} \Big( \mu_{1}\dot{\mathbf{H}}_{0}
    + \frac{8\pi}{3} \dot{\mathbf{M}} \Big)
    \label{m}
\end{equation}
is the magnetic moment of the particle induced by eddy currents. For illustration, in Fig.~\ref{fig1} we show the lines of the induced electric field inside the particle and the magnetic field of eddy currents.

\section{CLOSED LLG EQUATION FOR CONDUCTING NANOPARTICLES}
\label{Eff}

Now, we are ready to derive the closed LLG equation describing the magnetization dynamics in conducting nanoparticles. For this we need to calculate the average magnetic field $\overline{\mathbf{H}} = (1/V)\int_{V} \mathbf{H}_{1} d\mathbf{r}$ of eddy currents. Using Eq.~(\ref{H1}) and simple formulas
\begin{equation}
    \frac{1}{V} \int_{V} r^{2} d\mathbf{r} =
    \frac{3a^{2}}{5}, \quad
    \frac{1}{V} \int_{V} (\mathbf{m}\cdot
    \mathbf{r})\mathbf{r} d\mathbf{r} =
    \frac{a^{2}}{5}\mathbf{m},
    \nonumber
    \label{int}
\end{equation}
we obtain $\overline{\mathbf{H}} = (2\mu_{2}/ \mu_{1}a^{3}) \mathbf{m}$, and so the total effective magnetic field acting on the magnetization becomes
\begin{equation}
    \bm{\mathcal{H}}_{\mathrm{eff}} = \mathbf{H}_{
    \mathrm{eff}} - \tau_{\sigma} \dot{\mathbf{H}}_{0} -
    \frac{8\pi\tau_{\sigma}}{3\mu_{1}}\dot{\mathbf{M}}.
    \label{mean_H}
\end{equation}
With this result, the unclosed LLG equation (\ref{LLG}) reduces to the closed one
\begin{equation}
    \dot{\mathbf{M}} = -\gamma \mathbf{M} \times
    \tilde{\mathbf{H}}_{\mathrm{eff}}  + \frac{
    \tilde{\alpha}}{M}\, \mathbf{M} \times
    \dot{\mathbf{M}},
    \label{eff_LLG}
\end{equation}
where $\tilde{\mathbf{H}}_{\mathrm{eff}} = \mathbf{H}_{\mathrm{eff}} - \tau_{\sigma} \dot{\mathbf{H}}_{0}$, $\tilde{\alpha} = \alpha + \alpha_{\sigma}$, and
\begin{equation}
    \alpha_{\sigma} = \frac{8\pi \gamma M\tau_{
    \sigma}} {3\mu_{1}}.
    \label{relax}
\end{equation}

Thus, the magnetization dynamics in ferromagnetic metal particles can be described by the closed LLG equation (\ref{eff_LLG}). In this equation, which is the main result of this paper, the effects of particle conductivity are accounted by two terms. The first, $-\tau_{\sigma} \dot{\mathbf{H}}_{0}$, can be considered as an additional external magnetic field and the second, $\alpha_{\sigma}$, as an additional damping parameter. Both these terms arise from eddy currents inside the particle. However, while the first term results from eddy currents induced by changing the external magnetic field, the second term results from eddy currents induced by changing the magnetization direction. We note also that since $|\mathbf{H}_{1} |_{r=0}/ |\overline{\mathbf{H}}| = 2 + \mu_{1}/2\mu_{2}$, the exact result (\ref{relax}) is $5/2$ times less (at $\mu_{1} = \mu_{2} =1$) than that obtained in Ref.~[\onlinecite{MaLo}].

To clarify the importance of these terms in the magnetization dynamics, we first analyze the conditions under which Eq.~(\ref{eff_LLG}) is derived. In our model, we consider spherical ferromagnetic particles that are assumed to be single-domain. According to the Brown's fundamental theorem,\cite{Brow1} the single-domain state in these particles is energetically favorable if the particle radius $a$ is less than the critical radius $a_{\mathrm{cr}}$, which usually ranges from a few tens to a few hundreds of nanometers.

Next, the quasi-static approximation of Maxwell's equations implies\cite{LaLi2} that (a) the wavelength of the electromagnetic field is much larger than the particle size, (b) the displacement current $(1/4\pi) \partial \mathbf{E}_{1}/ \partial t$ is much smaller than the conduction current $\sigma \mathbf{E}_{1}$, and (c) the electric conductivity $\sigma$ and magnetic susceptibilities $\mu_{l}$ in the Maxwell equations (\ref{EH}) are the same as in the static case. Introducing the characteristic angular frequency $\omega$ of the electromagnetic field, the first two conditions can be written as $\omega \ll c/a$ and $\omega \ll \sigma$, respectively. The third condition for the conductivity is satisfied if the field period greatly exceeds the electron mean-free time $\tau_{0}$, i.e., if $\omega \tau_{0} \ll 1$. Since $\max{a} = a_{\mathrm{cr}}$, $\max{a_{\mathrm{ cr}}} \sim 10^{-5}\,  \mathrm{cm}$, for good conducting metals $\sigma \sim 10^{17} - 10^{18}\, \mathrm{s}^{-1}$, and $\tau_{0}$ at room temperature is of the order of $10^{-13}\, \mathrm{s}$, one can make sure that these three conditions of quasi-stationarity are equivalent to the single condition $\omega \tau_{0} \ll 1$. It should be noted, however, that the magnetic susceptibilities $\mu_{l}$ tend to $1$ as $\omega$ increases and the difference between $\mu_{l}$ and $1$ can vanish at $\omega \tau_{0} \ll 1$.\cite{LaLi2} In this case, it is necessary to replace $\mu_{l}$ by $1$ in all formulas obtained within the quasi-static approximation.

Finally, let us discuss the conditions under which the magnetostatic approximation can be used to determine the magnetic induction inside and outside the particle. It is clear from the previous analysis that this approximation is valid if $\mu_{l} |\mathbf{H}_{l}| \ll |\mathbf{B}_{l}|$. Using formulas (\ref{B1}) and (\ref{B2}) for the magnetic induction in the magnetostatic approximation and formulas (\ref{H1}) and (\ref{H2}) for the magnetic field of eddy currents, it can be easily verified that this inequality holds if $\omega \tau_{\sigma} \ll 1$. Hence, collecting all the conditions, we may conclude that the LLG equation (\ref{eff_LLG}) is valid if $a< a_{\mathrm{cr}}$ and $\omega \max\{ \tau_{\sigma}, \tau_{0} \} \ll 1$.

Since, according to Eqs.~(\ref{def_tau}) and (\ref{relax}), $\alpha_{\sigma} \sim \tau_{\sigma}$ and $ \tau_{\sigma} \sim a^{2}$, the influence of eddy currents on the magnetization dynamics increases with increasing the particle size and reaches the maximum at $a \sim a_{\mathrm{cr}}$. To estimate the parameters $\tau_{\sigma}$ and $\alpha_{\sigma}$ in this case, we assume that $a = 10^{-5}\, \mathrm{cm}$, $\sigma = 10^{18}\, \mathrm{s}^{-1}$, $4\pi M = 2 \times 10^{4}\, \mathrm{G} $, $\gamma = 1.76 \times 10^{7}\, \mathrm{s}^{-1} \mathrm{G} ^{-1}$, and $\mu_{1} = \mu_{2} = 1$. Then, from Eqs.~(\ref{def_tau}) and (\ref{relax}), one gets $\tau_{\sigma} \approx 9.31 \times 10^{-14}\, \mathrm{s}$ and $\alpha_{\sigma} \approx 2.18 \times 10^{-2}$. Due to the smallness of the characteristic time $\tau_{\sigma}$ and limitation of $\omega$, the term $-\tau_{\sigma} \dot{\mathbf{H} }_{0}$ can usually be neglected compared to the external magnetic field $\mathbf{H}_{0}$ and, as a consequence, the effective magnetic field $\tilde{\mathbf{H}}_{\mathrm{eff}}$ in Eq.~(\ref{eff_LLG}) can be replaced by $\mathbf{H}_{\mathrm{eff}}$. As to the additional damping parameter $\alpha_{\sigma}$, its influence on the magnetization dynamics is, in general, not negligible and it is most pronounced when $\alpha \lesssim \alpha_{\sigma}$.

To illustrate the role of $\alpha_{\sigma}$, we consider the behavior of the magnetization in the time-dependent magnetic field
\begin{equation}
    \mathbf{H}_{0}(t) = H_{0}\mathbf{e}_{x}
    \left\{\!\! \begin{array}{cl}
    t/\tau,
    & t \leq \tau
    \\ [4pt]
    1,
    & t > \tau.
    \end{array}
    \right.
    \label{H_0}
\end{equation}
Assuming that $\mathbf{e}_{a} = \mathbf{e}_{z}$, from Eqs.~(\ref{H_eff}) and (\ref{H^0_1}) at $\mu_{1} = \mu_{2} = 1$ we obtain
\begin{equation}
    \mathbf{H}_{\mathrm{eff}} = \frac{H_{a}}{M}(
    \mathbf{M} \cdot \mathbf{e}_{z})\mathbf{e}_{z}
    + \mathbf{H}_{0}(t).
    \label{H_eff2}
\end{equation}
[The demagnetization field $-(4\pi/3) \mathbf{M}$ in the right-hand side of Eq.~(\ref{H_eff2}) is omitted because it does not affect the magnetization dynamics.] Since the main features of the magnetization dynamics in the reference case has already been studied in the context of precessional switching of magnetization,\cite{BWH, BFH, KaRu, SMB, SuWa} here we only intend to show that the time behavior of $\mathbf{M}$ in dielectric and metallic nanoparticles can be qualitatively differen.

\begin{figure}
    \centering
    \includegraphics[totalheight=6cm]{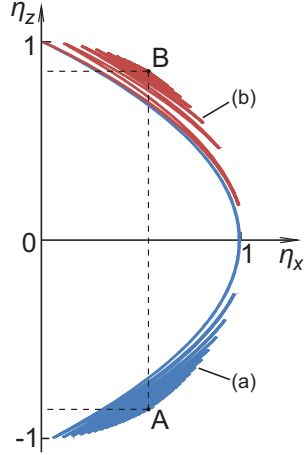}
    \caption{\label{fig2} (Color online) Trajectories of
    the reduced magnetization $\bm{\upeta} = \mathbf{M}/M$
    in the $(\eta_{x}, \eta_{z})$ plane. The time dependence
    of $\bm{\upeta}$ in dielectric and metallic nanoparticles
    is represented by the trajectories (a) and (b),
    respectively. The qualitative difference between these
    trajectories results solely from the action of the
    magnetic field of eddy currents.}
\end{figure}

Replacing $\tilde{\mathbf{H}}_{ \mathrm{eff}}$ by $\mathbf{H}_{ \mathrm{eff}}$, taking the cross product of both sides of Eq.~(\ref{eff_LLG}) with $\mathbf{M}$ and using the condition $\dot{\mathbf{M}} \cdot \mathbf{M} = 0$, the LLG equation (\ref{eff_LLG}) can easily be reduced to the LL equation
\begin{equation}
    \dot{\mathbf{M}} = -\frac{\gamma}{1 + \tilde{
    \alpha}^{2}} \mathbf{M} \times \mathbf{H}
    _{\mathrm{eff}} - \frac{\tilde{\alpha} \gamma}
    {(1 + \tilde{\alpha}^{2})M}\, \mathbf{M} \times
    (\mathbf{M} \times \mathbf{H}_{\mathrm{eff}}),
    \label{eff_LL}
\end{equation}
which is more convenient for numerical solution. Using the initial condition $\mathbf{M}(0) = M\mathbf{e}_{z}$, we solved this equation at $H_{a} = 5 \times 10^{3}\, \mathrm{Oe}$, $\tau = 10^{-12}\, \mathrm{s}$, $\alpha = 2 \times 10^{-2}$, $h = H_{0}/ H_{a} = 0.52$ and other parameters given above. The trajectories of $\mathbf{M}$ in the plane $(\eta_{x}, \eta_{z})$, where $\eta_{x,z} = M_{x,z}/M$, are shown in Fig.~\ref{fig2}. The magnetization dynamics for $\sigma=0$ is represented by the trajectory (a) that begins at the point with coordinates $\eta_{x} = 0$ and $\eta_{z} = 1$ (at $t=0$) and ends at the point A with coordinates $\eta_{x} = h$ and $\eta_{z} = -\sqrt{1 - h^{2}}$ (at $t = \infty$). Since $\eta_{z}$ at $t=0$ and $t=\infty$ has different signs, the magnetization switching occurs (the time at which $\eta_{z} =0$ approximately equals $7.29 \times 10^{-11} \, \mathrm{s}$). In contrast, the magnetization dynamics for $\sigma \neq 0$ is represented by the trajectory (b) that ends at the point B with coordinates $\eta_{x} = h$ and $\eta_{z} = \sqrt{1 - h^{2}}$, i.e., there is no magnetization switching in this case. This explicitly shows that eddy currents in ferromagnetic metal nanoparticles can significantly affect the magnetization dynamics.

\section{CONCLUSIONS}
\label{Concl}

We have developed an analytical model to describe the magnetization dynamics in ferromagnetic metal nanoparticles. It is based on the coupled system of the Landau-Lifshitz-Gilbert (LLG) equation for the magnetization, in which the effective magnetic field contains the magnetic field of eddy currents, and Maxwell's equations for the electromagnetic field induced by the external magnetic field and magnetization. We have analytically solved the Maxwell equations in the quasi-static approximation and have determined the magnetic field of eddy currents averaged over the particle volume. Using this result, we have derived the closed LLG equation describing the magnetization dynamics in metallic nanoparticles.

This equation contains two additional terms in comparison with the LLG equation that describes the magnetization dynamics in dielectric nanoparticles. The first term arises from eddy currents induced by changing the external magnetic field and is represented as an additional external magnetic field. In contrast, the second term results from eddy currents induced by changing the magnetization and is accounted for by an additional dumping parameter. We have shown that while the additional external magnetic field can be neglected in most cases, the additional dumped parameter may strongly influence the magnetization dynamics in relatively large metallic nanoparticles. This has been demonstrated by considering the precessional switching of magnetization in metallic and dielectric nanoparticles.

\section*{ACKNOWLEDGMENTS}

We are grateful to the Ministry of Education and Science of Ukraine for financial support under Grant No.~0112U001383.

\end{document}